\documentstyle[eqsecnum,aps,graphicx]{revtex}

\begin{document}
\draft

\title{The endpoint of the rp process on accreting neutron stars}

\author{H.~Schatz$^{1}$, A.~Aprahamian$^{2}$, V.~Barnard$^{2}$,
 L.~Bildsten$^{3}$,
A.~Cumming$^{3}$, M.~Ouellette$^{1}$, T.~Rauscher$^{4}$, F.-K.~Thielemann$^{4}$,
M.~Wiescher$^{2}$}

\address{$^{1}$ Dept.\ of Physics and Astronomy and National 
Superconducting Cyclotron Laboratory, 
Michigan State University, East Lansing, MI 48824}
\address{$^{2}$ Dept.\ of Physics, University of Notre Dame, Notre Dame, 
IN 46556}
\address{$^{3}$ Institute for Theoretical Physics, University of California,
Santa Barbara, CA 93106}
\address{$^{4}$ Dept.\ of Physics and Astronomy, Universit\"at Basel, Klingelbergstr. 82, CH-4056, Basel, Switzerland}

\date{\today}
\maketitle

\begin{abstract}
We calculate the rapid proton (rp) capture process of hydrogen burning
on the surface of an accreting neutron star with an updated reaction
network that extends up to Xe, far beyond previous work. In both
steady-state nuclear burning appropriate for rapidly accreting neutron
stars (such as the magnetic polar caps of accreting X-ray pulsars) and
unstable burning of Type I X-ray bursts, we find that the rp process
ends in a closed SnSbTe cycle. This prevents the synthesis of
elements heavier than Te and has important consequences for X-ray
burst profiles, the composition of accreting neutron stars, and
potentially galactic nucleosynthesis of light p nuclei.
\end{abstract}



\section{Introduction}

 Observations of X-ray binaries are an important source of
information about neutron stars. In these systems a neutron star accretes
hydrogen and helium rich matter from the envelope of a close companion
star. Gravitational energy is released when the matter reaches the
neutron star and powers most of the observable X-ray radiation.  The
accreted material is compressed, heated, and eventually undergoes
thermonuclear burning. The released nuclear energy, though small
relative to the gravitational energy, is observable as Type I X-ray
bursts at accretion rates $\lesssim 10^{-8} M_\odot \ {\rm yr^{-1}}$.
For these rates, the nuclear burning proceeds unstably 
\cite{WoT76,MaC77,Jos77,Lew95}, repeating on typical timescales of hours or days.
In bright X-ray sources and in magnetically accreting X-ray pulsars
the accretion rates reach or locally exceed the Eddington limit. The
accreted layer is then thermally stable and the thermonuclear burning
proceeds in steady state \cite{JoL80,SBC98}.

When hydrogen is present, the nuclear burning proceeds via the rapid
proton capture process (rp process) \cite{WaW81}, a sequence of proton
captures and $\beta$ decays responsible for the burning of hydrogen
into heavier elements.  We show here, for the first time, that there
is a natural termination of this process due to a closed cycle in the
Sn-Te region (SnSbTe cycle).

Understanding the rp process and especially 
its endpoint is a prerequisite for the
interpretation of X-ray burst light curves, which show a wide variety
of characteristics including millisecond oscillations
\cite{SSZ98}. Furthermore, the ashes of the nuclear burning replace
the original crust of the neutron star and determine its thermal and
electrical conductivities. The nuclear burning products thus  have 
direct repercussions for the evolution of magnetic fields
\cite{Brb98}, the quiescent luminosity in transient X-ray binaries
\cite{BBR98} (which could be used to discriminate against systems with
accreting black holes), and the emission of potentially detectable
gravitational waves from a rapidly rotating neutron star with a
deformed crust \cite{Bil98,UBC00}. There has also been an interest in
the nuclear burning processes during X-ray bursts as a potential site
for the nucleosynthesis of some light p nuclei like $^{92,94}$Mo and
$^{96,98}$Ru that are severely underproduced in standard p process
scenarios (for example \cite{RAH95}). This would require the escape of
some of the burned material into the interstellar medium.

 Many early rp process simulations for X-ray bursts were based on
reaction networks that ended at $^{56}$Ni. For these calculations, it
was argued that further proton captures were negligible for hydrogen
consumption and energy production.  
However, several later 
studies were performed with larger networks ending in the
Kr-Y region \cite{WaW81,FSL87,HSH83,WGI94,KHA99} or using a
simplified 16 nuclei network up to Cd \cite{WaW84}. Recently
a parameter study was carried out with an updated and
extended network up to Sn \cite{SAG97}.
All of these studies
found that a significant fraction of the rp process reached the end of
their respective networks.
An rp process reaching 
the end of a reaction network up to Sn was also clearly demonstrated
for the steady state burning regime for accretion rates between 20 and
40 times the Eddington limit ($\dot{m}_{\rm
Edd}=$8.8~10$^4$~g/cm$^2$/s) \cite{SBC98}.

\section{Calculations}

 In this paper we explore the rp process beyond Sn in explosive and
steady state burning to determine for the first time the endpoint 
of the rp process. Our one-zone X-ray burst model is based on the
physics outlined in reference \cite{Bil97}.  Realistic ignition
conditions have been calculated as a function of accretion rate and
metallicity \cite{CuB00}.  For the X-ray burst model we assume an
accretion rate of 0.1~$\dot{m}_{\rm Edd}$, a flux out of the crust of
0.15~MeV per accreted nucleon, and a metallicity $Z=10^{-3}$.  An
accreted layer with solar metallicity would produce a similar burst at
a higher accretion rate of 0.3~$\dot{m}_{\rm Edd}$.  The steady state
model is the same as described in reference \cite{SBC98}. For all
models we assume accretion of hydrogen and helium in solar
proportions, and a surface gravity of 1.9$\times 10^{14}$cm/s$^2$.

The nuclear reaction network includes all proton rich nuclei from
hydrogen to xenon and was updated relative to the data described
in \cite{SAG97}. The theoretical Hauser-Feshbach reaction rates
have been recalculated with
the new Hauser-Feshbach code NON-SMOKER \cite{RaT00}.
A more detailed discussion of the nuclear
physics input and the X-ray burst model
will be published in a forthcoming paper.

\section{Results}

\begin{figure}[ht]
\includegraphics[width=6.5in]{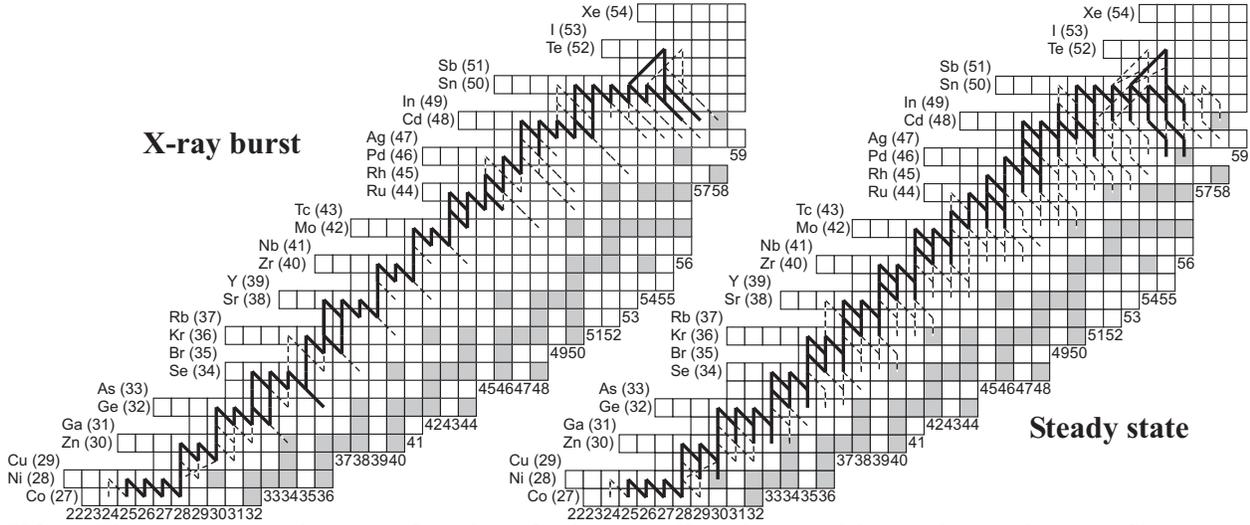}
\caption{The time integrated reaction flow above Ga during an X-ray burst
and for steady state burning. Shown are reaction flows of more than
10\% (solid line) and of 1-10\% (dashed line) of the reaction flow
through the 3$\alpha$ reaction.
\label{FigFlow}}
\end{figure}

Figure~\ref{FigFlow} shows our results for
the time integrated reaction flow during an
X-ray burst.  Ignition takes place at a density of
$1.1\times 10^6$~g/cm$^3$ and the burst reaches a peak temperature of
1.9~GK, with a rise time scale of $\approx$4~s, and a cooling phase
lasting $\approx$200~s. Helium burns via the
3$\alpha$ reaction, and the $\alpha$p process \cite{WaW81}, a sequence
of alternating ($\alpha$,p) and (p,$\gamma$) reactions into the Sc
region. These helium burning processes provide the seed nuclei for the
rp process. The
rp process reaction flow reaches the Sn isotopes
in the $^{99}$Sn-$^{101}$Sn range $\approx$ 80~s (time for half maximum)
after the burst peak
and proceeds then along the Sn isotopic
chain towards more stable nuclei.

\begin{figure}[ht]
\includegraphics[width=2.6in]{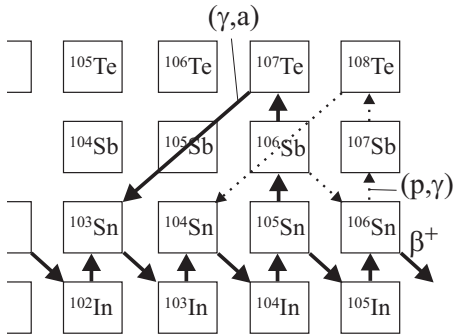}
\caption{The reactions in the SnSbTe cycles during an X-ray burst.
In the case of proton captures the arrows indicate the direction 
of the net flow, the difference of the flow via proton capture
and the reverse flow via ($\gamma$,p) photodisintegration. The
line styles are the same as in figure \protect\ref{FigFlow}.
\label{FigCycle}}
\end{figure}

Processing beyond Sn occurs if the corresponding Sb isotone
is sufficiently proton bound for the ($\gamma$,p) photodisintegration to
be small. This occurs at $^{105}$Sn. However, after two proton
captures a strong $^{107}$Te($\gamma$,$\alpha$) photodisintegration
rate cycles the reaction flow back to $^{103}$Sn. The reaction path is characterized by a
cyclic flow pattern, the SnSbTe cycle which represents the endpoint for
the rp-process reaction flow towards higher masses (see figure \ref{FigCycle}). 
The SnSbTe cycle forms because
the neutron deficient $^{106-108}$Te isotopes are $\alpha$ unbound by
$\approx$4~MeV. In fact, $^{107}$Te is a known ground state $\alpha$
emitter \cite{SKK79}.  A fraction of the reaction flow proceeds via $\beta$ decay
of $^{105}$Sn into $^{106}$Sn, and the reaction sequence
$^{106}$Sn(p,$\gamma$)$^{107}$Sb(p,$\gamma$)$^{108}$Te($\gamma$,$\alpha$)$^{104}$Sn
leads to a second, weaker cycle.
Calculations with different ignition conditions confirm that the
rp process cannot proceed beyond the SnSbTe cycles.

A previous calculation of the rp process in steady state burning
found that most material accumulated at the end of the network
(the Sn isotopes) for an accretion rate of $40 \dot{m}_{\rm Edd}$ 
\cite{SBC98}.
Figure~\ref{FigFlow} shows the reaction flow at that accretion rate. We
find that the rp process ends in a similar SnSbTe cycle as in X-ray
bursts. Some of the material is now cycled back via
$^{106}$Sb(p,$\alpha$), which successfully competes with $^{106}$Sb(p,$\gamma$) at
steady state burning conditions.  Calculations at different accretion rates
show that the rp process can never overcome the closed SnSbTe cycle.
For steady state burning, we are
now able to compute accurately the composition of the ashes for all
accretion rates.

\begin{figure}[ht]
\includegraphics[width=2.6in]{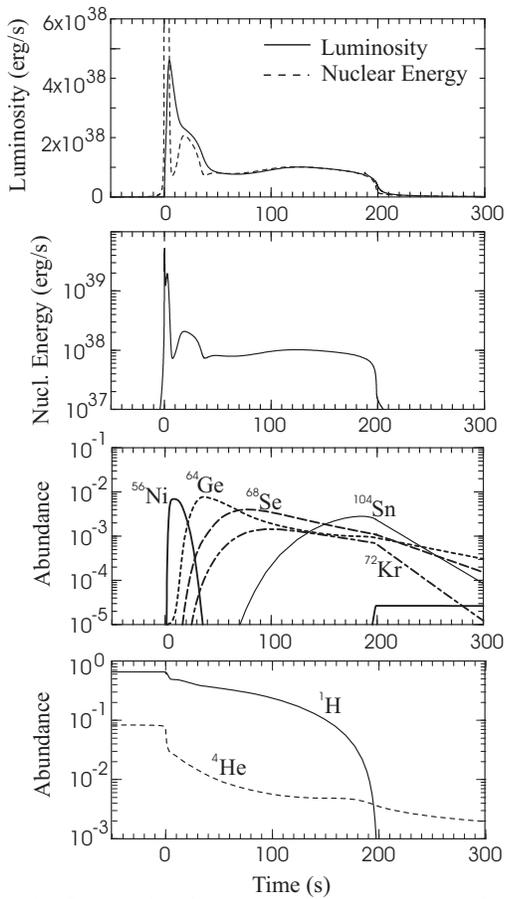}
\caption{Luminosity, nuclear energy generation rate, 
and the abundances of hydrogen, helium and the important waiting point
nuclei as functions of time
during an X-ray burst. For comparison, the nuclear energy generation 
rate is also shown as a dashed line together with the luminosity,
though it is out of scale during the peak of the burst. 
The mass of the accreted layer is 4.9$\times 10^{21}$~g.
\label{FigTime}}
\end{figure}

The SnSbTe cycle impacts the light-curve of X-ray bursts and
the consumption of hydrogen. This is illustrated in figure~\ref{FigTime},
which shows the correlation between the X-ray burst luminosity,
the abundances of some important long lived nuclei (waiting points)
in the rp process,
and the hydrogen and helium abundances. Clearly the slow hydrogen burning
via the rp process
beyond $^{56}$Ni is responsible for the extended burst tail.
The SnSbTe cycle
builds up the abundance of the longest lived nucleus in the cycle,
$^{104}$Sn (20.8~s half-life), and produces helium towards the
end of the burst.  This triggers an increase in the 3$\alpha$ flow
and subsequently an increase in energy production and hydrogen consumption. As a
consequence, the burst lasts longer and hydrogen is completely
burned.

\begin{figure}[ht]
\includegraphics[width=3in]{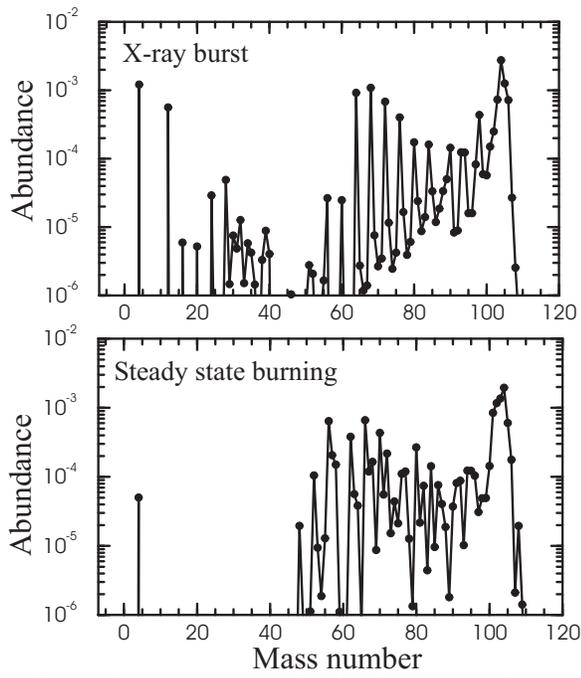}
\caption{Final abundance distribution as functions of mass number for
an X-ray burst, and for steady state burning at an accretion rate
of 40$\dot{m}_{\rm Edd}$.
\label{FigFinab}}
\end{figure}

 The SnSbTe cycle also affects the composition of the rp process
ashes, shown for the X-ray burst and the steady state calculation in
figure~\ref{FigFinab}. The limitation imposed on the rp process by
the SnSbTe cycle is clearly reflected in the
lack of nuclei heavier than $A=107$.
Nevertheless we obtain a broad distribution of
nuclei in the $A=64$-107 mass range. This is a result of the long
lived waiting point nuclei
along the rp process reaction path which store
some material until the burning is over. The late helium production in
the SnSbTe cycle broadens this distribution further.

\section{Conclusions}
  To summarize, we have shown that the synthesis of heavy nuclei via
the rp process is limited to nuclei with $Z\le 54$ due to our newly
discovered SnSbTe cycle. The existence of a SnSbTe cycle under all rp
process conditions is a consequence of the low, experimentally known
\cite{PWC94} $\alpha$ separation energies of the
$^{106,107,108,109}$Te isotopes and is therefore not subject to
nuclear physics uncertainties. However, because of the uncertainties
in the proton separation energies of the Sb isotopes there is some
uncertainty in the relative strength of the SnSbTe sub-cycles closed
by ($\gamma,\alpha$) photodisintegration on $^{106}$Te, $^{107}$Te,
and $^{108}$Te. This will be discussed in a forthcoming paper.

A likely consequence of the SnSbTe cycle for accreting neutron stars
is that the matter entering the crust is composed of
nuclei lighter than $A\approx$107.  The only way to bypass the SnSbTe
cycle would be a pulsed rp process, where between pulses matter could
decay back to stable nuclei.  This could happen during so called dwarf
bursts, which have been suggested to be secondary bursts produced by
reignition of the ashes \cite{TWW93}. However, this would require some
unburned hydrogen in the burst ashes (see discussion below) or
extensive vertical mixing \cite{FSL87}.

  Our calculations give a strong indication that the synthesis of
nuclei beyond $^{56}$Ni and especially into the $A=100$ mass region in
hydrogen rich bursts leads to extended energy production. This might
explain the long duration (100 seconds) bursts seen from, for example,
GS 1826-24 \cite{KHK00}.

The importance of the endpoint of the rp process for the consumption
of hydrogen in X-ray bursts has been discussed extensively. Previous
calculations limited to nuclei up to $^{56}$Ni always found unburned
hydrogen. Later burning of this hydrogen was proposed to explain short
burst intervals \cite{TWW93}, occasionally observed extended X-ray
flares \cite{TWL96},
and spin down in
millisecond oscillations, as recently discovered during the tail of
a long X-ray burst from 4U~1636-53 \cite{Str99}.
It was pointed out before that depending on
the endpoint of the rp process these models might largely overestimate
the amount of unburned hydrogen \cite{FSL87}. Our calculations indicate
indeed that the SnSbTe cycle reduces the amount of unburned hydrogen
substantially. Within the framework of the one-zone model we find that 
all the hydrogen is burned, a full multizone hydrodynamic burst
calculation is needed to confirm this result. 

  In our X-ray burst calculation, we find large overproduction factors
(produced abundance to solar abundance $\approx 10^9$) for the p
nuclei $^{98}$Ru, $^{102}$Pd, and $^{106}$Cd. Overproduction factors
of this order have been suggested to be sufficient to explain the
origin of the solar system abundances of these nuclei if $\approx 1$\%
of the burned material is somehow ejected \cite{SAG97}. However, in
order to accurately calculate the composition of the ejected material
the ejection mechanism has to be identified.  Furthermore we find that
the synthesis of p nuclei is most likely associated with X-ray bursts
having long tails.  As these bursts are rare, we conclude that only a
small fraction of X-ray bursts are likely to produce light p nuclei in
the $A=90-106$ mass range. Therefore, considerably larger amounts of
matter would have to be ejected in X-ray bursts to account for the
galactic abundances of light p nuclei than previously
assumed. Nevertheless it is striking that the SnSbTe cycle happens to
naturally limit rp process nucleosynthesis to light p nuclei.

 The authors would like to thank R. Boyd and E. Roeckl for helpful
discussions. Part of this work was carried out under NSF contract
PHY-95-28844, PHY-99-01133 and PHY-99-07949, under NASA grant
NAG5-8658, and under and
the Swiss NSF contracts 2124-055832.98 and 2000-061822.00.
L.B. is a Cottrell Scholar of the Research Corporation.
V.B. was supported through the Graduate School of the
University of Notre Dame.


\newcommand{\noopsort}[1]{} \newcommand{\printfirst}[2]{#1}
  \newcommand{\singleletter}[1]{#1} \newcommand{\swithchargs}[2]{#2#1}

\newpage

\end{document}